\documentclass[twocolumn,showpacs,preprintnumbers,amsmath,amssymb]{revtex4}
\usepackage{graphicx}
\usepackage{dcolumn}
\usepackage{bm}

\begin{document}

\title{Short-Time Decoherence for General System-Environment Interactions}

\author{Denis Tolkunov}\email{tolkunov@clarkson.edu}
\author{Vladimir Privman}\email{privman@clarkson.edu}
\affiliation{Center for Quantum Device Technology, Department of
Physics, Clarkson University, Potsdam, New York 13699-5721}


\begin{abstract}
Short time approximation is developed for system-environmental bath mode
interactions involving a general  non-Hermitian system operator $\Lambda$,
and its conjugate, $\Lambda^\dagger$, in order to
evaluate onset of decoherence at low temperatures in quantum
systems interacting with environment. The developed approach is
complementary to Markovian approximations and appropriate for
evaluation of quantum computing schemes. Example of a spin system
coupled to a bosonic heat bath via $\Lambda \propto \sigma_-$ is
worked out in detail.
\end{abstract}


\maketitle

\section{\label{sec:level1}Introduction}

Quantum system exposed to environmental modes is described by the
reduced density matrix, and its evolution deviates from the ideal,
usually pure-state, dynamics. For short times, appropriate for
quantum computing gate functions and, generally, for controlled
quantum dynamics, approximation schemes for the density matrix
have been developed recently
[\onlinecite{Loss}-\onlinecite{Onset2}]. The present work derives
a new rather general short-time approximation which applies for
models with system-bath interactions involving a general system
operator. It thus extends the previously known approach
[\onlinecite{Onset1},\onlinecite{Onset2}] which was limited to
couplings involving a single Hermitian system operator.

We consider an open quantum system with the Hamiltonian
\begin{gather}
H=H_S+H_B+H_I.  \label{eq:HHH}
\end{gather}
Here $H_S$ describes the system proper. It is coupled to the
environment (bath), described by $H_B$. The system and bath are coupled by the
interaction $H_I$. The bath has been traditionally modelled
[\onlinecite{Loss}-\onlinecite{Legg}] by a large number of
uncoupled bosonic modes, namely harmonic oscillators (with ground
state energy shifted to zero),
\begin{gather}
H_B=\sum\limits_k \omega _k^{\vphantom{\dagger}} b_k^{\dagger
}b_k^{\vphantom{\dagger}}.
\end{gather}
Here $b_k$ are the bosonic annihilation operators corresponding to
the bath modes, and from now on we use the convention $\hbar=1 $.

In most of this work, we consider the general system $H_S$, and we
assume that the interaction with the bath involves the system
operator $\Lambda$ that couples linearly
[\onlinecite{VKam}-\onlinecite{Grab}] to the bath modes,
\begin{gather}
H_I = \Lambda \sum\limits_k g_k^{\vphantom{\dagger}} b_k^{\dagger
} + \Lambda ^{\dagger } \sum\limits_k g_k^{*}
b_k^{\vphantom{\dagger}},  \label{eq:one}
\end{gather}
with the interaction constants $g_k$.

Let $R(t)$ denote the overall density matrix. It is commonly
assumed [\onlinecite{Loss}-\onlinecite{Grab}] that at time $t=0$
the system and bath are not entangled, and the bath modes are
thermalized,
\begin{gather}\label{RR}
R\left( 0\right) =\rho \left( 0\right) \prod\limits_k\theta _k,
\end{gather}
where
\begin{gather}\label{theta}
\theta _k={\rm Z}_k^{-1}e^{-\beta \omega
_k^{\vphantom{\dagger}}b_k^{\dagger }b_k^{\vphantom{\dagger}}},
\end{gather}
with $ \beta=1 /kT $, and
\begin{gather}
{\rm Z}_k\equiv\left( 1-e^{-\beta \omega _k}\right) ^{-1}.
\end{gather}

We point out that while the quantum system $S$, described by the
reduced density matrix $\rho(t)$, is small, typically
two-state (qubit) or several-qubit, the bath has many degrees of
freedom. The combined effect of the bath modes on the system can
be large even if each of them is influenced little by the system.
This has been the basis for the arguments for the harmonic
approximation for the bath modes
[\onlinecite{Loss}-\onlinecite{Legg}] and the linearity of the
interaction, as well as for the Markovian approximations
[\onlinecite{VKam}-\onlinecite{Grab}] that assume that the bath
modes are ``reset'' to the thermal state by the ``rest of the
universe'' on time scales shorter than any dynamical time of the
system interacting with the bath.

The frequencies of the oscillators of the bath are usually assumed
to be distributed from zero to some cutoff value $\omega_c$. The
bath modes with the frequencies close to the energy gaps of the
system, $\Delta E_{ij}=E_i-E_j$, contribute to the ``resonant''
thermalization and decoherence processes. Within the Markovian
schemes, the diagonal elements of the reduced density matrix of the system,
\begin{equation}
\rho \left( t\right) =\mathrm{Tr}_B R\left( t\right) ,
\end{equation}
approach the thermal values $\propto e^{-E_i/kT} $ for large times
exponentially, on time scale $T_1$. The off-diagonal elements
vanish, which represents decoherence, on time scale $T_2$,
which, for resonant processes, is
given by $T_2 \simeq 2T_1$.
However, generally decoherence is expected to be faster than
thermalization because, in addition to resonant processes, it can
involve virtual processes that do not conserve energy. It has been
argued that this additional ``pure'' decoherence is dominated by
the bath modes with near-zero frequencies
[\onlinecite{VKam},\onlinecite{Grab},\onlinecite{VKam2}]. At low
temperatures, this ``pure decoherence'' is expected
[\onlinecite{12}] to make $T_2 \ll T_1$.

Since the resetting of these low-frequency modes to the thermal
state occurs on time scales $1/kT=\beta$, the Markovian approach
cannot be used at low temperatures
[\onlinecite{VKam},\onlinecite{Grab},\onlinecite{VKam2}].
Specifically, for quantum computing in solid-state
semiconductor-heterostructure architectures
[\onlinecite{12}-\onlinecite{22}], temperatures as
low as few $10${\,}mK are needed. This brings the thermal time
scale to $\beta \sim 10^{-9}\,$sec, which is close to the
 single-qubit control times $10^{-11} \hbox{-} 10^{-7}\,$sec
[\onlinecite{12}-\onlinecite{22}]. Alternatives to
the Markovian approximation have been suggested
[\onlinecite{Ford}-\onlinecite{Scnon}].

In this work, we generalize the recently suggested scheme
[\onlinecite{Onset1},\onlinecite{Onset2}], applicable for
Hermitian $\Lambda$ only, to a wider class of interaction
Hamiltonians. We treat the case when the system operator $\Lambda$
entering the interaction, see (\ref{eq:one}), is not Hermitian. In
actual applications in quantum computing, calculations with only a
single qubit or few qubits are necessary for evaluation of the
local ``noise,'' to use the criteria for quantum error correction
[\onlinecite{Shor}-\onlinecite{Presk}]. For example, the system
Hamiltonian is frequently taken proportional to the Pauli matrix
$\sigma_z$. The interaction operator $\Lambda$ can be proportional
to $\sigma_x$, which is Hermitian. Such cases are covered by the
short-time approximation developed earlier
[\onlinecite{Onset1},\onlinecite{Onset2}]. However, one can also
consider models with $\Lambda \propto \sigma_-$. Similarly, models
with non-Hermitian $\Lambda$ are encountered in Quantum Optics
[\onlinecite{Lois}]. In Section II, we develop our short time
approximation scheme. Results for a spin-boson type model are
given in Section III.

\section{Short Time Approximation}

In this section we obtain a general expression for the time
evolution operator of the system (\ref{eq:HHH}-\ref{eq:one})
within the short time approximation. The system operators $H_S$
and $\Lambda$ need not be specified at this stage; the derivation
is quite general.

In order to define ``short time,'' we consider dimensionless
combinations involving the time variable $t$. There are several
time scales in the problem. These include the inverse of the
cutoff frequency of the bath modes, $1 / {\omega _c}$, the thermal
time $\beta= 1 / {kT}$, and the internal characteristic times of
the system $1 / {\Delta E_{ij}}$. Also, there are time scales
associated with the system-bath interaction-generated
thermalization and decoherence, $T_{1,2}$. The shortest time scale
at low temperatures (when $\beta$ is large) is typically $1 /
{\omega _c}$. The most straightforward expansion in $t$ yields a
series in powers of ${\omega _c} t $. The aim of developing more
sophisticated short-time approximations
[\onlinecite{Onset1},\onlinecite{Onset2}] is to preserve unitarity
and obtain expressions approximately valid up to intermediate
times, of order of the system and interaction-generated time
scales. The latter property can only be argued for heuristically
in most cases, and checked by model calculations.

The overall density matrix, assuming time-independent Hamiltonian
over the quantum-computation gate function time intervals
[\onlinecite{12}-\onlinecite{22}], evolves
according to
\begin{gather}
R\left( t\right) =U(t)R\left( 0\right) [U(t)]^{\dagger },
\end{gather}
where
\begin{gather}
U(t)=e^{-i\left( H_S+H_B+H_I\right) t}  \label{eq:evolut}
\end{gather}
is the evolution operator.

The general idea of our approach is the following. We break the
exponential operator in (\ref{eq:evolut}) into products of simpler
exponentials. This involves an approximation, but allows us to
replace system operators by their eigenvalues, when spectral
representations are used, and then calculate the trace of
$R(t)$ over the bath modes, obtaining explicit expressions for
the elements of the reduced density matrix of the system.
For Hermitian coupling operators, $\Lambda^{\dagger}=\Lambda$,
our approach reduces to known results [\onlinecite{Onset1},\onlinecite{Onset2}].

We split the exponential evolution operator into terms that do not have
any noncommuting system operators in them. This requires an approximation.
For short times, we start by using the factorization
[\onlinecite{Kirz}-\onlinecite{Sorn}]
\begin{eqnarray}
&&e^{-i\left( H_S+H_B+H_I\right) t+O\left( t^3\right) }  \nonumber
\label{HB} \\
&=&e^{-\frac i2H_St}e^{-i\left( H_I+H_B\right) t}e^{-\frac
i2H_St},
\end{eqnarray}
where we have neglected terms of the third and higher orders in $t$,
in the exponent. The
middle exponential in (\ref{HB}),
\begin{equation}
\Xi \equiv e^{-i\left( H_B+H_I\right) t}=e^{-i\left( H_B+\Lambda G^{\dagger }+\Lambda ^{\dagger
}G\right) t},
\end{equation}
where
\begin{equation}
G\equiv \sum\limits_kg_k^{*}b_k,
\end{equation}
still involves noncommuting terms as long as $\Lambda$ is non-Hermitian. In terms of
the Hermitian operators
\begin{eqnarray}
&&L\equiv {\frac 12}\left( \Lambda +\Lambda ^{\dagger }\right) ,  \label{lam1} \\
&&M\equiv {\frac i2}\left( \Lambda -\Lambda ^{\dagger }\right) ,
\label{lam2}
\end{eqnarray}
we have
\begin{gather}
\Lambda G^{\dagger }+\Lambda ^{\dagger }G=L\left( G+G^{\dagger
}\right) +iM\left( G-G^{\dagger }\right) .
\end{gather}
We then carry out two additional short-time factorizations within the same quadratic-in-$t$
(in the exponent) order of
approximation,
\begin{eqnarray}\label{eq:decomp3}
\Xi &= & e^{\frac 12\left[ M\left(
G-G^{\dagger }\right) -iH_B\right] t}e^{\frac i2H_Bt}\\\nonumber
&\times&e^{-i\left[ H_B+L\left( G+G^{\dagger }\right) \right]
t}e^{\frac i2H_Bt}e^{\frac 12\left[ M\left( G-G^{\dagger }\right)
-iH_B\right] t}.
\end{eqnarray}
This factorization is chosen in such a way that $\Xi$ remains unitary,
and for $M=0$ or $L=0$ the expression is identical to that used for the
Hermitian case [\onlinecite{Onset1},\onlinecite{Onset2}]. The evolution operator then takes the form
\begin{gather}
U=e^{-\frac i2H_St}\,\Xi \,e^{-\frac i2H_St},  \label{uu}
\end{gather}
with $\Xi$ from (\ref{eq:decomp3}), which is an approximation in terms of a product of several
unitary operators.

It has been recognized [\onlinecite{Onset1},\onlinecite{Onset2}]
that approximations of this sort are superior to the
straightforward expansion in powers of $t$ (or more exactly,
$\omega_c t$). Specifically, in (\ref{HB}), we notice that $H_S$ is
factored out in such a way that $H_B$, which commutes with $H_S$,
droppes out of all the commutators that enter the higher-order
correction terms. This suggests that a redefinition of the
energies of the modes of $H_B$ should have only a limited effect
on the corrections and serves as a heuristic argument for the
approximation being valid beyond the shortest time scale
$1/\omega_c$, up to intermediate time scales.

Our goal is to approximate the reduced density matrix of the
system. We consider its energy-basis matrix elements,
\begin{equation}
\rho _{mn}\left( t\right) =\mathrm{Tr}_B\left\langle m\right|
UR\left( 0\right) U^{\dagger }\left| n\right\rangle ,  \label{rho}
\end{equation}
where
\begin{gather}
H_S\left| n\right\rangle =E_n\left| n\right\rangle .  \label{eig}
\end{gather}
We next use the factorization (\ref{HB},\ref{eq:decomp3}) to systematically replace
system operators by c-numbers, by inserting decompositions of the
unit operator in the bases defined by $H_S$, $L$, and $M$. First,
we collect the expressions (\ref{RR},\ref{eq:decomp3},\ref{uu},
\ref{eig}), and use two energy-basis decompositions of unity to
get
\begin{eqnarray}\label{NS}
\rho _{mn}\left( t\right) &=&\sum\limits_{p\,q}e^{\frac i2\left(
E_n+E_q-E_m-E_p\right) t}\rho _{pq}\left( 0\right)  \nonumber   \\
&\times& \mathrm{Tr}_B\left[ \left\langle m\right| \Xi \left|
p\right\rangle \right. \prod\limits_k\theta _k\left. \left\langle
q\right| \Xi ^{\dagger }\left| n\right\rangle \right] .
\end{eqnarray}

We now define the eigenstates of $L$ and $M$,
\begin{gather}
L\left| \lambda \right\rangle =\lambda \left| \lambda
\right\rangle ,
\label{eigen1} \\
M\left| \mu \right\rangle =\mu \left| \mu \right\rangle .
\label{eigen2}
\end{gather}
The operators $\Xi $ and $\Xi ^{\dagger }$ introduce exponentials
in (\ref{NS}) that contain either $L$ or $M$ in the power. By
appropriately inserting $\sum\limits_\lambda \left| \lambda
\right\rangle \left\langle \lambda \right| $ or $\sum\limits_\mu
\left| \mu \right\rangle \left\langle \mu \right| $ between these
exponentials, we can convert all the remaining system operators to
c-numbers.
For convenience, let us define the operators
\begin{eqnarray}
\pi _{\alpha \beta \gamma }=\left| \alpha \right\rangle
\left\langle \alpha \right| \left. \beta \right\rangle
\left\langle \beta \right. \left| \gamma \right\rangle
\left\langle \gamma \right| \,\,\;\;\;\;\;
\end{eqnarray}
and
\begin{eqnarray}\label{U_op}
{\cal U}_{s_1,s_2,s_3}=e^{s_1^{\vphantom{\dagger}}g_{k^{\vphantom{,}}}^{*}
b_k^{\vphantom{\dagger}}t+s_2^{\vphantom{\dagger}}g_k^{\vphantom{\dagger}}b_k^{\dagger
}t-is_3^{\vphantom{\dagger}}\omega
_k^{\vphantom{\dagger}}b_k^{\dagger }b_k^{\vphantom{\dagger}}t}.
\end{eqnarray}

\

\noindent{}The resulting expression for the trace entering (\ref{NS}) is
\begin{eqnarray}\label{SS}
&&\mathrm{Tr}_B\left\{ \left\langle m\right| \Xi \left|
p\right\rangle \right. \prod\limits_k\theta _k\left. \left\langle
q\right| \Xi ^{\dagger }\left| n\right\rangle \right\}\\\nonumber
=&&\sum\limits_{\mu _j\lambda _j}\left\langle m\right| \pi _{\mu
_1\lambda _1\mu _2}\left| p\right\rangle \left\langle q\right| \pi
_{\mu _3\lambda _2\mu _4}\left| n\right\rangle
\prod\limits_k{\cal T}_k,
\end{eqnarray}
where the indices $\lambda$ and $\mu$ label the eigenstates of $L$ and $M$, respectively, and
\begin{eqnarray}\label{trk}
{{\cal T}_k}&=&\mathrm{Tr}_k\left\{ {\cal U}_{\frac 12\mu _1,-\frac 12\mu
_1,\frac 12}{\cal U}_{0,0,-\frac 12}{\cal U}_{-i\lambda _1,-i\lambda
_1,1}\right.\\&\times& {\cal U}_{0,0,-\frac 12}{\cal U}_{\frac 12\mu _2,-\frac
12\mu _2,\frac 12}\theta _k {\cal U}_{-\frac 12\mu _3,\frac 12\mu
_3,-\frac 12} \nonumber   \\\nonumber &\times& \left. {\cal U}_{0,0,\frac
12}{\cal U}_{i\lambda _2,i\lambda _2,-1}{\cal U}_{0,0,\frac 12}{\cal U}_{-\frac 12\mu
_4,\frac 12\mu _4,-\frac 12}\right\} .
\end{eqnarray}

In order to calculate the trace over the $k$th bath mode in
(\ref{trk}), we rearrange the operators using the cyclic property,
in such a way that the formula (\ref {identity}), derived in
Appendix A, can be used to simplify products of two or three
operators $\cal U$ at a time. For example, we can transfer the operator
${\cal U}_{\frac 12\mu _1,-\frac 12\mu _1,\frac 12}$ to the right
hand side, getting the combination
\begin{equation}
{\cal U}_{-\frac 12\mu _4,\frac 12\mu _4,-\frac 12}{\cal U}_{\frac 12\mu
_1,-\frac 12\mu _1,\frac 12},
\end{equation}
inside the trace, and use the identity (\ref{identity}). We then
transfer ${\cal U}_{0,0,-\frac 12}$ to the right hand side and
repeat the process, now for the three rightmost operators. After several steps we arrive to the following
expression for the trace,
\begin{equation}
{{\cal T}_k}={\cal R}_k\mathrm{Tr}_k\left\{ \theta _k {\cal U}_{u,-u^*,0}\right\} ,
\label{ftr}
\end{equation}
where
\begin{eqnarray}  \label{PQ}
u &=&\frac i{\omega _kt}\left( e^{-\frac{i\omega _kt}2}-1\right)
\mu _{-}^{\prime \prime }-\frac{2i}{\omega _kt}e^{-\frac{i\omega
_kt}2}\sin
\frac{\omega _kt}2\lambda _{-}  \nonumber \\
&+&\frac i{\omega _kt}e^{-\frac{i\omega _kt}2}\left(
e^{-\frac{i\omega _kt} 2}-1\right) \mu _{-}^{\prime }
\end{eqnarray}
and
\begin{widetext}
\begin{eqnarray}
{\cal R}_k =\exp \Big\{\!\! &-&\! \frac{i\left| g_k\right|
^2}{\omega _k^2}\Big[ \big( \sin \frac{\omega _kt}2-\frac{\omega
_kt}2\big) \left( \mu _{-}^{\prime }\mu _{+}^{\prime }+\mu
_{-}^{\prime \prime }\mu _{+}^{\prime \prime }\right) +\left( \sin
\omega _kt-\omega _kt\right) \lambda _{-}\lambda
_{+}       \nonumber  \label{RK} \\
&-& 4\sin \frac{\omega _kt}2\sin ^2\frac{\omega _kt}4\mu
_{-}^{\prime }\mu _{+}^{\prime \prime } -2 \sin ^2\frac{\omega
_kt}2\left( \lambda _{-}\mu _{+}^{\prime \prime }-\lambda _{+}\mu
_{-}^{\prime }\right)  \Big] \Big\} .
\end{eqnarray}
\end{widetext}
Here we introduced the variables
\begin{eqnarray}
\mu _{\pm }^{\prime } &=&\mu _1\pm \mu _4 , \\
\mu _{\pm }^{\prime \prime } &=&\mu _2\pm \mu _3,
\end{eqnarray}
and
\begin{equation}
\lambda _{\pm }=\lambda _1\pm \lambda _2.
\end{equation}
The trace in (\ref{ftr}) can be evaluated, for instance, by using the
coherent states technique, see Appendix \ref{ap2},
\begin{gather}
{{\cal T}_k}={\cal R}_ke^{\frac{pq}2\left| g_k\right| ^2t^2\coth
\frac{\beta \omega _k}2}.  \label{TK}
\end{gather}
The expression which follows from (\ref{SS}), (\ref{PQ}),
(\ref{RK}) and (\ref{TK}) is
\begin{gather}
\prod\limits_k{{\cal T}_k}=\exp \left[ -\mathcal{P}\left( t\right)
\right] , \label{DME1}
\end{gather}
where
\begin{widetext}
\begin{eqnarray}
\mathcal{P} &=&B^2\left( t\right) \left( \lambda _{-}^2+\mu
_{-}^{\prime }\mu _{-}^{\prime \prime }\right) +B^2\left(
t/2\right) \left( \mu _{-}^{\prime \prime }-\mu _{-}^{\prime
}\right) ^2-F\left( t\right) \left(
\mu _{-}^{\prime \prime }-\mu _{-}^{\prime }\right) \lambda _{-} \\
&-&iC\left( t\right) \lambda _{-}\lambda _{+}-iC\left( t/2\right)
\left( \mu _{-}^{\prime }\mu _{+}^{\prime }+\mu _{-}^{\prime
\prime }\mu _{+}^{\prime \prime }\right) +iS\left( t\right) \left(
\lambda _{-}\mu _{+}^{\prime \prime }-\lambda _{+}\mu _{-}^{\prime
}\right) -iC_1\left( t\right) \mu _{-}^{\prime }\mu _{+}^{\prime
\prime }.  \nonumber
\end{eqnarray}
\end{widetext}
The coefficients here are the spectral sums over the
bath modes,
\begin{eqnarray}\label{spec1}
B^2\left( t\right) =2\sum\limits_k\frac{\left| g_k\right|
^2}{\omega _k^2} \sin ^2\frac{\omega _kt}2\coth \frac{\beta \omega
_k}2,
\end{eqnarray}
\begin{eqnarray}\label{spec2}
C\left( t\right) =\sum\limits_k\frac{\left| g_k\right| ^2}{\omega
_k^2} \left( \omega _kt-\sin \omega _kt\right) ;
\end{eqnarray}
these functions are well known [\onlinecite{35},\onlinecite{27}]. The
result also involves the new spectral functions
\begin{eqnarray}\label{spec3}
S\left( t\right) =-2\sum\limits_k\frac{\left| g_k\right|
^2}{\omega _k^2}\sin ^2\frac{\omega _kt}2,
\end{eqnarray}
\begin{eqnarray}\label{spec4}
F\left( t\right) =4\sum\limits_k\frac{\left| g_k\right| ^2}{\omega
_k^2}\sin ^2\frac{\omega _kt}4\sin \frac{\omega _kt}2\coth
\frac{\beta \omega _k}2.
\end{eqnarray}
Furthermore, for the sake of convenience we defined
\begin{equation}\label{C1}
C_1\left( t\right)
=2C\left( t/2\right) -C\left( t\right) .
\end{equation}

By using (\ref{NS}) and (\ref{DME1}), we obtain our final result for the density matrix
evolution,
\begin{eqnarray}  \label{eq:EV}
\rho _{mn}\left( t\right) &=&\sum\limits_{p,q}\sum\limits_{\mu
_j\lambda _j}e^{\frac i2\left( E_n+E_q-E_m-E_p\right) t}\rho
_{pq}\left( 0\right)\\\nonumber &\times& \left\langle m\right| \pi
_{\mu _1\lambda _1\mu _2}\left| p\right\rangle \left\langle
q\right| \pi _{\mu _3\lambda _2\mu _4}\left| n\right\rangle
e^{-\mathcal{P}} , \nonumber
\end{eqnarray}
where the first sum over $p$ and $q$ is over the energy
eigenstates of the system; the second sum is over $\lambda_1
,\lambda_2$ and $\mu_1 , \ldots , \mu_4$, which label the
eigenstates of the operators $L$ and $M$, respectively; see
(\ref{eigen1}) and (\ref{eigen2}).

\section{Discussion and Application}

The result (\ref{eq:EV}) looks formidable in the general case.
However, in most applications evaluation of decoherence will
require short-time expressions for the reduced density matrix of a
single qubit. Few- and multi-qubit systems will have to be treated
by utilizing additive quantities
[\onlinecite{norm}-\onlinecite{addnorm}], accounting for quantum error
correction (requiring measurement), etc. For a two-state
system---a qubit---the summation in (\ref{eq:EV}) involves
$2^8=64$ terms, each a product of several factors calculation of
which is straightforward. Still, the required bookkeeping is
cumbersome, and we utilized the symbolic language Mathematica to carry
out the calculation for an illustrative example.

We consider the model [\onlinecite{Swain}] defined by
\begin{gather}
H=\mathcal{A}{\,}\sigma _z+\sum\limits_k\omega
_k^{\vphantom{\dagger} }b_k^{\dagger
}b_k^{\vphantom{\dagger}}+\sum\limits_k\left(
g_k^{\vphantom{\dagger}}\sigma _{-}b_k^{\dagger }+g_k^{*}\sigma
_{+}b_k^{\vphantom{\dagger}}\right) ,  \label{RWA}
\end{gather}
where $\mathcal{A} \geq 0$ is a constant, $\sigma _{\pm}={1\over 2}
(\sigma_x \pm \sigma_y)$ and $\sigma _z$ are the Pauli matrices, $b_k^{\dagger }$ and
$b_k^{\vphantom{\dagger}}$ are the bosonic creation and
annihilation operators, and $g_k$ are the coupling constants.
Physically this model may describe, for example, a qubit
interacting with a bath of phonons, or a two-level molecule in an
electromagnetic field. In the latter case, this is a variant of
the multi-mode Jaynes-Cummings model
[\onlinecite{Lois},\onlinecite{JC}]. Certain spectral properties
of this model, the
field-theoretic counterpart of which is known as the Lee field
theory, are known analytically, e.g., [\onlinecite{Pf}]. However,
the trace over the bosonic modes, to obtain the reduced density
matrix for the spin, has not been obtained exactly.

For the model (\ref{RWA}) we have $\Lambda =\sigma _{-}$ and
$\Lambda ^{\dagger }=\sigma _{+}$, so that $L=\sigma _x/2$ and
$M=\sigma _y/2.$
We have $\left| \lambda _{1,2 }\right\rangle =\left( \left|
\uparrow \right\rangle \pm \left| \downarrow \right\rangle \right)
/\sqrt{2}$, with eigenvalues $ \lambda _{1,2 }=\pm 1/2$, and
$\left| \mu _{1,2 }\right\rangle =\left(\left| \uparrow
\right\rangle \pm i\left| \downarrow \right\rangle \right) /
\sqrt{2}$, with eigenvalues $\mu _{1,2 }=\pm 1/2$.
For the initial state, let us assume that the spin at $t=0$ is in
the excited state $\left| \uparrow \right\rangle \left\langle
\uparrow \right| $, so that the initial density matrix has the
form
\begin{gather}
\rho \left( 0\right) =\left(
\begin{array}{cc}
1 & 0 \\
0 & 0
\end{array}
\right) .
\end{gather}
Calculation in
Mathematica yields the following results for the density matrix
elements, $\rho_{12}(t)=0$ and
\begin{widetext}
\begin{eqnarray}\label{Density}
4\rho _{11}\left( t\right)&=&2+\,e^{-2B^2\left( t\right) }+
e^{-4B^2\left( \frac t2\right) }\cosh \left( 2\,F\right)
+2e^{-2B^2\left( \frac t2\right) }\sinh \left( B_1\right) \cos
\left( S\right) \\ &+& 2e^{-B^2\left( \frac t2\right) }\cos \left(
C_1\right) \sin \left( S\right)   \nonumber +ie^{-B^2\left(
t\right) -B^2\left( \frac t2\right) }\left[ \,e^{iC_1}\sinh \left(
-iS+F\right) +\,e^{-iC_1}\sinh \left( -iS-F\right) \right],
\nonumber
\end{eqnarray}
\end{widetext}
where $C_1$ was defined in (\ref{C1}) and
\begin{equation}
B_1(t)=2B^2\left( t/2\right) -B^2\left( t \right) .
\end{equation}
Where not explicitly shown,
the argument of all the spectral functions entering
(\ref{Density}) is $t$.

In order to obtain irreversible behavior and evaluate a measure of
decoherence, we consider the continuum limit of infinitely many
bath modes. We introduce the density of the bosonic bath states
$\cal{D}\left(\omega\right)$, incorporating a large-frequency
cutoff $\omega_c$, and replace the summations in
(\ref{spec1})-(\ref{spec4})  by integrations over $\omega$
[\onlinecite{Legg},\onlinecite{VKam},\onlinecite{35},\onlinecite{52}].
For instance, (\ref{spec1}) takes the form,
\begin{eqnarray}
\ B^2\left( t\right) =\int\limits_0^\infty d\omega
\frac{{\cal{D}}\left( \omega \right) |g( \omega )|^2
}{\omega ^2}\sin ^2\frac{\omega t}2\coth \frac{\beta \omega
}2   .
\end{eqnarray}
We will use the standard
Ohmic-dissipation [\onlinecite{Legg}] expression, with an exponential cutoff, for
an illustrative calculation,
\begin{eqnarray}
{\cal{D}}\left( \omega \right) |g( \omega )|^2 =\Omega \,
\omega  \, e^{-\omega /\omega_c},
\end{eqnarray}
where $\Omega $ is a constant.

We point out that the results obtained for
the density matrix elements depend on the dimensionless variable
$\omega_c t$, as well as on the dimensionless parameters $\Omega$
and $\omega_c \beta$ ($= \hbar \omega_c / kT$, where we remind the
reader that $\hbar$, set to 1, must be restored in the final
results). Interestingly, the results do not depend explicitly on
the energy gap parameter $\mathcal{A}$, see (\ref{RWA}). This
illustrates the point that short-time approximations do not
capture the ``resonant'' relaxation processes, but rather only
account for ``virtual'' relaxation/decoherence processes dominated
by the low-frequency bath modes. However, the short-time approximations
of the type considered here are meaningful only for systems with
well-defined separation of the resonant vs.\ virtual decoherence
processes, i.e., for $\hbar / \mathcal{A} \gg 1/\omega_c$. For such
systems, $\hbar / \mathcal{A} = 1 / \mathcal{A}$ defines one of the ``intermediate''
time scales beyond which the approximation cannot be trusted.

\begin{figure}[tbp]
\includegraphics[width=9cm, height=9cm]{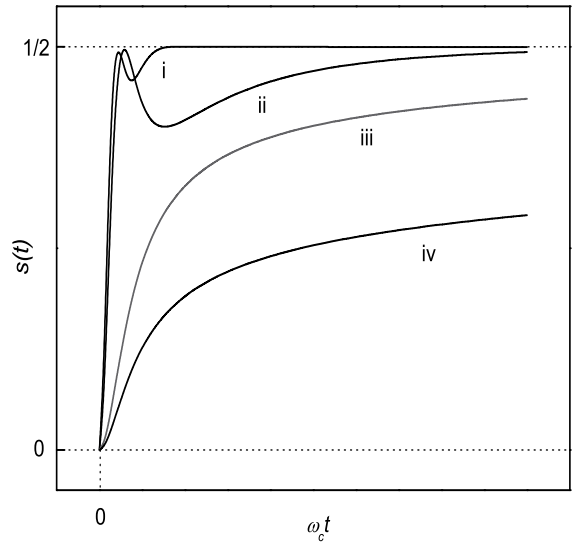}
\caption{Schematic behavior of $s\left(t\right)$ for different
values of  $\Omega$, decreasing from $\sf{i}$ to $\sf{iv}$. }
\label{graph1}
\end{figure}

As an example, we calculated a measure of deviation of a qubit
from a pure state in terms of the ``linear entropy''
[\onlinecite{norm},\onlinecite{addnorm},\onlinecite{24}],
\begin{gather}
s(t)=1-\mathrm{Tr}{\,}\left[\rho ^2\left( t\right)\right].
\end{gather}
Figure\ 1 schematically illustrates the behavior of $s(t)$ for
different $\Omega$ values, for the case $\omega_c^{-1} <<  \beta
$. The values of $s(t)$ increase from zero, corresponding to a
pure state, to $1/2$, corresponding to a completely mixed state,
with superimposed oscillations. For Ohmic dissipation, three time
regimes can be identified [\onlinecite{27}]. The shortest time
scale is set by $t<O\left(1/{\omega _c}\right) $. The
quantum-fluctuation dominated regime corresponds to
$O\left(1/{\omega _c}\right) <t<O\left( 1/{kT}\right) $. The
thermal-fluctuation dominated regime is $t>O\left( 1/{kT}\right)
$. Our short time approximation yields reasonable results in the
first two regimes. For $t>O\left( 1/{kT}\right) $ it cannot correctly
reproduce the process of thermalization. Instead, it predicts
approach to the maximally mixed state.

\begin{figure}[tbp]
\includegraphics[width=9cm, height=9cm]{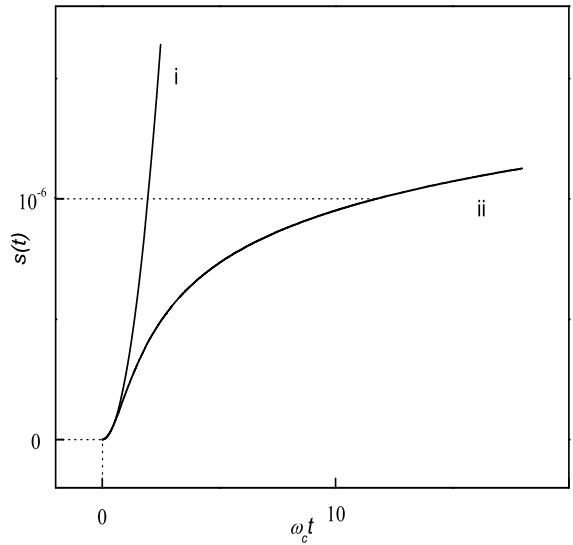}
\caption{The comparison between the $O(t^2)$ expansion, $\sf{i}$, and the
short-time approximation, $\sf{ii}$.} \label{graph2}
\end{figure}

Figure\ 2 corresponds to the parameter values typical for low
temperatures and appropriate for quantum computing applications,
$\omega_c \beta=10^3$, with $\Omega=1.5\cdot 10^{-7}$ chosen to
represent weak enough coupling to the bath to have the decoherence
measure reach the threshold for fault-tolerance, of order
$10^{-6}$, for ``gate'' times well exceeding $1/ \omega_c$, here
for $ \omega_c t $ over 10. The leading-order quadratic expansion
in powers of the time variable $t$ is also shown. Its validity is
limited to $t<O\left(1/{\omega _c}\right) $ and it cannot be used
for evaluation of quantum-computing models.

This research was supported by the National Security Agency
and Advanced Research and Development Activity under Army Research
Office contract DAAD-19-02-1-0035, and by the
National Science Foundation, grant DMR-0121146.

\appendix

\section{}

Our aim is to derive a relation of the form
\begin{equation}  \label{identity}
{\cal U}_{v_1,w_1,x}{\cal U}_{v_3,w_3,0}{\cal U}_{v_2,w_2,-x}=\kappa{\cal U}_{p,q,0},
\end{equation}
where the operator ${\cal U}_{s_1,s_2,s_3}$ was defined in (\ref{U_op}).
Consider the quantity
\begin{equation}
\Delta =e^{x\left( b^{\dagger }+\alpha _1\right) \left( b+\alpha
_2\right) }e^{\gamma _1b^{\dagger }+\gamma _2b}e^{-x\left(
b^{\dagger }+\beta _1\right) \left( b+\beta _2\right) },
\label{del111}
\end{equation}
where $b^{\dagger }$, $b$ are the bosonic creation and annihilation
operators, $x$, $\alpha _i$, $\beta _i$ are c-numbers. Let us use
the identity [\onlinecite{Lois}],
\begin{gather}
e^{\alpha b-\beta b^{\dagger }}f\left( b,b^{\dagger }\right)
e^{-\alpha b+\beta b^{\dagger }}=f\left( b+\beta ,b^{\dagger
}+\alpha \right) , \label{shift}
\end{gather}
to represent the first and third exponentials in $\Delta $ in
the form
\begin{eqnarray}
e^{x\left( b^{\dagger }+\alpha _1\right) \left( b+\alpha _2\right)
}&=&e^{\alpha _1b-\alpha _2b^{\dagger }}e^{xb^{\dagger
}b}e^{-\alpha
_1b+\alpha _2b^{\dagger }},  \label{e1} \\
e^{-x\left( b^{\dagger }+\beta _1\right) \left( b+\beta _2\right)
}&=&e^{\beta _1b-\beta _2b^{\dagger }}e^{-xb^{\dagger }b}e^{-\beta
_1b+\beta _2b^{\dagger }}.  \label{e2}
\end{eqnarray}
We then combine the second exponential in (\ref{del111}) and
the last and first exponentials in (\ref{e1}) and (\ref{e2}),
by utilizing the identity
\begin{gather}
e^{\alpha b}e^{\beta b^{\dagger }}=e^{\frac 12\alpha \beta
}e^{\alpha b+\beta b^{\dagger }},  \label{bh}
\end{gather}
which follows from (\ref{shift}). The resulting exponential
operator, with exponent linear in $b$ and $b^{\dagger }$, is
sandwiched between $e^{xb^{\dagger }b}$ and $e^{-xb^{\dagger }b}$.
Therefore, the following identity can be utilized
[\onlinecite{Lois}],
\begin{gather}
e^{xb^{\dagger }b}f\left( b,b^{\dagger }\right) e^{-xb^{\dagger
}b}=f\left( be^{-x},b^{\dagger }e^x\right) .
\end{gather}
Once again using (\ref{bh}), we arrive at the
following expression,
\begin{eqnarray}  \label{del}
\Delta =e^{\nu b+\mu b^{\dagger }+r},
\end{eqnarray}
where
\begin{eqnarray}
&&\mu =\left( \alpha _2-\beta _2\right) \left( e^x-1\right) +\gamma _1e^x, \\
&&\nu =\left( \beta _1-\alpha _1\right) \left( e^{-x}-1\right)
+\gamma _2e^{-x},
\end{eqnarray}
and\begin{widetext}
\begin{eqnarray}\nonumber
r &=&-2\left( \alpha _1\beta _2-\alpha _2\beta _1\right) \sinh
^2\frac x2+\left( \alpha _1\alpha _2-\beta _1\beta _2\right) \sinh
x \\  &+&\frac 12\gamma _1\left( \alpha _1+\beta _1\right)
\left( e^x-1\right) +\frac 12\gamma _2\left( \alpha _2+\beta
_2\right) \left( e^{-x}-1\right) .
\end{eqnarray}
Now (\ref{identity}) follows, with
\begin{eqnarray}
\kappa &=&\exp \left[ \frac{2\left| g_k\right|
^2}{x^2\omega _k^2}\sin ^2\left(
\frac{x\omega _kt}2\right) \left( v_1w_2-v_2w_1\right) \right]  \nonumber \\
&\times& \exp \left[ \frac{i\left| g_k\right| ^2}{x^2\omega
_k^2}\left( \sin \left( x\omega _kt\right) \left(
v_1w_1-v_2w_2\right) +x\omega _kt\left(
v_2w_2-v_1w_1\right) \right) \right] \nonumber \\
&\times& \exp \left[ \frac{i\left| g_k\right| ^2t}{2x\omega
_k}\left( \left( e^{-ix\omega _kt}-1\right) w_3\left(
v_1-v_2\right) +\left( e^{ix\omega _kt}-1\right) v_3\left(
w_1-w_2\right) \right) \right],
\end{eqnarray}
\end{widetext}
and
\begin{eqnarray}
p =-\frac i{x\omega _kt}\left( e^{ix\omega _kt}-1\right) \left(
v_1+v_2\right) +v_3e^{ix\omega _kt}\,\,\,\,\,\, , \\
q =\frac i{x\omega _kt}\left( e^{-ix\omega _kt}-1\right) \left(
w_1+w_2\right) +w_3e^{-ix\omega _kt}.
\end{eqnarray}

\section{}\label{ap2}

Let us calculate the trace in (\ref{ftr}) which has the form
\begin{gather}\label{tr1}
{\cal T}\equiv \mathrm{Tr}\left\{ e^{\delta b^{\dagger
}b}e^{vb+wb^{\dagger }}\right\} ,
\end{gather}
where we omitted the index $k$ since all the calculations here are
in the space of a single mode. We use the coherent-state technique
[\onlinecite{Lois}]. The coherent states $\left| z\right\rangle $
by definition are eigenstates of the annihilation operator, $b$
\begin{gather}
b\left| z\right\rangle =z\left| z\right\rangle ,
\end{gather}
with complex eigenvalues $z=x+iy$. These states are not orthogonal
\begin{gather}
\left\langle z_1\right| \left. z_2\right\rangle =\exp \left(
z_1^{*}z_2-\frac 12\left| z_1\right| ^2-\frac 12\left| z_2\right|
^2\right) ,
\end{gather}
and they form an overcomplete set. The identity operator
can be written as
\begin{gather}
\int d^2z\left| z\right\rangle \left\langle z\right| =1,
\end{gather}
where the integration in complex plane is defined via
\begin{gather}
d^2z=\frac 1\pi dxdy.
\end{gather}

We represent the trace (\ref{tr1}) by the coherent-state
integral using the relation
\begin{gather}
\mathrm{Tr}A=\int d^2z\left\langle z\right| A\left|
z\right\rangle,
\end{gather}
where $A$ is an arbitrary operator. We then use the normal
ordering, $\mathcal{N}$, formula for bosonic operators, represented
schematically (see [\onlinecite{Lois}] for details) by
\begin{gather}\label{normal}
e^{\delta b^{\dagger }b}=\mathcal{N}e^{b^{\dagger }\left( e^\delta
-1\right) b}.
\end{gather}
The second term in the trace in (\ref{tr1}) is split by using
(\ref{bh}). All instances of $b$ and $b^\dagger$ can then be
replaced by $z$ and $z^*$, and the integral evaluated to yield the
expression for the trace,
\begin{gather}
{\cal T}=\frac {e^{\frac{wv}2\coth \frac \delta 2}}{1-e^\delta }.
\end{gather}

\end{document}